\documentclass[aps,secnumarabic,nobalancelastpage,
nofootinbib,onecolumn,12pt,tightenlines]{revtex4}%
\usepackage{amsmath}
\usepackage{graphics}
\usepackage{graphicx}
\usepackage{longtable}
\usepackage{url}
\usepackage{bm}
\usepackage{amsfonts}
\usepackage{amssymb}%
\providecommand{\U}[1]{\protect\rule{.1in}{.1in}}
\begin{document}
\title{Poisson-Nernst-Planck-Fermi Theory for Ion Channels }
\author{Jinn-Liang Liu}
\affiliation{Department of Applied Mathematics, National Hsinchu University of Education,
Hsinchu 300, Taiwan. E-mail: jinnliu@mail.nhcue.edu.tw}
\author{Bob Eisenberg}
\affiliation{Department of Molecular Biophysics and Physiology, Rush University, Chicago,
IL 60612 USA. E-mail: beisenbe@rush.edu}
\thanks{A different version of this paper was published in J. Chem. Phys. 141 (2014) 22D532.}
\date{\today}

\begin{abstract}

\end{abstract}
\maketitle

\textbf{Abstract.} A Poisson-Nernst-Planck-Fermi (PNPF) theory is developed
for studying ionic transport through biological ion channels. Our goal is to
deal with the finite size of particle using a Fermi like distribution without
calculating the forces between the particles, because they are both expensive
and tricky to compute. We include the steric effect of ions and water
molecules with nonuniform sizes and interstitial voids, the correlation effect
of crowded ions with different valences, and the screening effect of water
molecules in an inhomogeneous aqueous electrolyte. Including the finite volume
of water and the voids between particles is an important new part of the
theory presented here. Fermi like distributions of all particle species are
derived from the volume exclusion of classical particles. Volume exclusion and
the resulting saturation phenomena are especially important to describe the
binding and permeation mechanisms of ions in a narrow channel pore. The Gibbs
free energy of the Fermi distribution reduces to that of a Boltzmann
distribution when these effects are not considered. The classical Gibbs
entropy is extended to a new entropy form --- called Gibbs-Fermi entropy ---
that describes mixing configurations of all finite size particles and voids in
a thermodynamic system where microstates do not have equal probabilities. The
PNPF model describes the dynamic flow of ions, water molecules, as well as
voids with electric fields and protein charges. The model also provides a
quantitative mean-field description of the charge/space competition mechanism
of particles within the highly charged and crowded channel pore. The PNPF
results are in good accord with experimental currents recorded in a 10$^{8}%
$-fold range of Ca$^{2+}$ concentrations. The results illustrate the anomalous
mole fraction effect, a signature of L-type calcium channels. Moreover,
numerical results concerning water density, dielectric permittivity, void
volume, and steric energy provide useful details to study a variety of
physical mechanisms ranging from binding, to permeation, blocking,
flexibility, and charge/space competition of the channel.

\section{Introduction}

Biological functions of proteins depend on the details of the mixtures of
ionic solutions found outside and inside cells. Trace concentrations (%
$<$
$10^{-6}$ M) of calcium ions (Ca$^{2+}$) and other signaling molecules provide
physiological control of many biological pathways and proteins inside cells
\cite{E13a}. For example, voltage-gated calcium (Ca$_{\text{V}}$) channels
exhibit the anomalous mole fraction effect that effectively blocks abundant
monovalent cations by a trace concentration of Ca$^{2+}$ ions
\cite{AM84,FT89,RN02}. The fundamental mechanism of the calcium channel is of
great technological and biological interest \cite{SM03,TC14}. Multiscale
analysis seems necessary since calibrated all atom simulations of trace
concentrations of ions in physiological solutions are not likely to be
available in the near future.

Interactions between diffusion and migration in the electric field are central
to the biologists' view of channels \cite{H51,H01}. Following the
drift-diffusion (DD) model in semiconductors, Eisenberg et al.
\cite{CB92,EC93,EK95} have proposed the Poisson-Nernst-Planck (PNP) model to
calculate rather than assume \cite{H51,H01} the electric field and then the
ionic current in biological ion channels. Interactions of ions and flows in
narrow channels, and saturation in binding sites are also central to the
biologists' view of channels \cite{H51,H01}. It has been difficult to combine
the two views --- computed electric fields in channels showing interactions
and saturation --- because charges in PNP (and quasiparticles in DD) are
points with no diameter and so cannot saturate the aqueous channel \cite{E13a}
the way real ions do. Recently, the saturation of spheres of any size has been
described by a Fermi like distribution derived in \cite{LE13} from the
configuration entropy of mixtures of ions of any diameter and composition. The
steric effect has been shown to be very important to adequately describe
equilibrium systems \cite{LE13,L13,LE14}.

We extend the Poisson-Fermi model \cite{LE13} in two important rather novel
ways. We include the excluded volume of water molecules and the `empty space'
created by packing constraints and voids between particles. The equilibrium
model is also generalized here to a nonequilibrium model called the
Poisson-Nernst-Planck-Fermi (PNPF) model that can describe flow, including the
steric effect of all particles, the correlation effect of ions and water
molecules, the screening effect of water, as well as the charge/space
mechanism in the channel pore at and away from equilibrium. This treatment
unites diffusion and electric current, with interactions and binding in narrow channels.

Both discrete and continuum forms of Gibbs free energy of the electrolyte are
developed in this paper. The Gibbs-Fermi free energy functional for the Fermi
distribution is shown to reduce to the Gibbs-Boltzmann functional for the
classical Boltzmann distribution when both steric and correlation effects are
not present. Moreover, a new entropy form called the Gibbs-Fermi entropy is
proposed here to connect the spatial distribution of ions, water, and voids
between them (that may vary) with the change of local probabilities of each
species (and void volume) which of course usually have different sizes. The
Gibbs-Fermi entropy is a consistent generalization of the global Boltzmann
entropy and the classical local Gibbs entropy widely used to describe systems
without steric and packing constraints.

\textbf{\underline{\textbf{The steric effect.}}} The steric effect of crowding
produces a steric energy term in PNPF that is a quantitative statement of the
crowded charge effect of charge/space competition. The charge/space
competition theory introduced by Nonner and Eisenberg to explain calcium
selectivity has been developed in a long series of papers using Monte Carlo
methods by Boda and Henderson, and density functional methods by Gillespie and
collaborators. This `all spheres' approach successfully describes almost all
selectivity properties of calcium channels and the main properties of sodium
channels such as the micromolar Ca$^{2+}$ affinity for L-type calcium channels
\cite{BV07,GB08,BV09}, the wide range of Ca$^{2+}$ affinities for different
types of calcium channels, and the switch in selectivity from calcium to
sodium when the side chains of the selectivity filter are switched from EEEE
(glu glu glu glu) to DEKA (asp glu lys ala) \cite{BV07,MG09}. It also accounts
for the selectivity between monovalent cations of different size
\cite{BH01,BB02,GX05} and for the self-organized pore structures for selective
ions \cite{GF11}. In the biologically crucial and special case of Na$^{+}$ vs
K$^{+}$ selectivity in the DEKA sodium channel (so central to the function of
the nervous system and metabolic budget of mammals with large brains
\cite{AR09,M09}), control variables can even be identified that
\textit{independently} control selectivity and binding \cite{BV07,MG09}.

\textbf{\underline{\textbf{Interactions.}}} Our main goal is to show how
interactions of diffusion, electrophoretic migration, steric exclusion, and
imperfect packing of particles can all be treated quantitatively in a unified
framework to analyze binding and flow in crowded ion channels, without
explicitly calculating forces between individual ions, water, or voids. We
show that a Fermi like distribution is able to describe these interactions
well enough to account for a wide range of important properties of ions in
channels. We wonder how well this approach can describe the myriad of nonideal
properties reported in the physical chemistry experimental literature (for
more than a century) which have escaped canonical description up to now
\cite{K38,ZC86,P95,JP00,LM03,F04,LT07,LJ08,KF09,HR11}.

Numerical results produced by the PNPF model are in accord with the
experimental data reported by Almers and McCleskey in 1984 for Ca$_{\text{V}}
$ channels over a 10$^{8}$-fold range of concentrations of calcium ions
\cite{AM84}. Their experimental data has been a benchmark for selectivity ever
since. Their data has been used as a target for models using a variety of
methods ranging from physiological and crystallographic \cite{TC14}, to
molecular dynamics (MD) \cite{LF01,BG02}, Brownian dynamics (BD)
\cite{CK00,CA01,KL13}, Monte Carlo (MC) \cite{BB00,BV07,BV09}, as well as
continuum approaches \cite{NE98,NC00,HL12}.

The remaining part of the paper is organized as follows. A derivation of the
configuration entropy of all hard-sphere ions and water with voids is proposed
in Section 2, where a Fermi type of excess chemical potential, Gibbs-Fermi
free energy functional, and Gibbs-Fermi entropy are also introduced. All these
models seem to be new to the literature, as far as we know, because they treat
finite size water molecules and voids explicitly. In Section 3, we extend the
Gibbs-Fermi theory to the Poisson-Nernst-Planck-Fermi theory for studying
ionic transport, steric energy, water density, and void distribution in
equilibrium or nonequilibrium conditions. In Section 4, a molecular-continuum
model specific to L-type calcium channels is presented to show how to
implement in a consistent way the PNPF theory of the molecular filter region
of few particles joined to the bath region of numerous particles. Section 5
demonstrates that PNPF currents agree quite well with the experimental
currents reported in \cite{AM84} under the same membrane potential and the
same 10$^{8}$-fold range of Ca$^{2+}$ concentrations measured in the
experiment. These conductance results seem to fit data better than other
models we know of. Some concluding remarks are given in Section 6.

\section{Fermi Distribution and Gibbs-Fermi Entropy}

Based on the configurational entropy model proposed in \cite{LE13} for aqueous
electrolytes with arbitrary $K$ species of nonuniform size, hard spherical
ions, we extend the free energy of the model to
\begin{equation}
F(N)=\phi\sum_{j=1}^{K+1}q_{j}N_{j}-k_{\text{B}}T\ln W \tag{1}%
\end{equation}
by including specifically the excluded volume effect of the next species
($K+1$) of water molecules. Here, $\phi$ is the electrostatic potential,
$N_{j}$ is the total number of $j$ species particles carrying the charge
$q_{j}$ $=z_{j}e$ with the valence $z_{j}$, $e$ is the proton charge, $k_{B}$
is the Boltzmann constant, and $T$ is the absolute temperature. The volume of
a $j$ type particle is $v_{j}=4\pi a_{j}^{3}/3$ with radius $a_{j}$. It is
important to note that water is treated as a polarizable hard sphere with zero
net charge in Eq. (1), so $z_{K+1}=q_{K+1}=0$. The polarizability of water and
the inclusion of voids represent important generalizations from the classical
primitive solvent model used to describe calcium channels \cite{NG01}. The
last term in (1) describes the mixing entropy of all ions and water molecules
over a total of $N$ available nonuniform sites in a system with
\begin{equation}
W=%
{\displaystyle\prod\limits_{j=1}^{K+1}}
W_{j}=\frac{N!}{\left(  \Pi_{j=1}^{K+1}N_{j}!\right)  \left(  N-\sum
_{j=1}^{K+1}N_{j}\right)  !}, \tag{2}%
\end{equation}
where $W_{1}=N!/(N_{1}!(N-N_{1})!)$ is the number of combinations for the
distribution of $N_{1}$ in all vacant sites $N$. $W_{2}=(N-N_{1}%
)!/(N_{2}!(N-N_{1}-N_{2})!)$ is the number of combinations for the
distribution of $N_{2}$ in $N-N_{1}$ vacant sites after $N_{1}$ being
distributed, and so on. After all particles are distributed, there remains (in
this model) just a single site $N_{K+2}=N-\sum_{j=1}^{K+1}N_{j}=1$ that is
used to represent the (continuously connected) voids created by defects in the
packing structure of all particles of all types and by Coulomb and steric
forces (e.g., Lennard-Jones) between particles. This void structure is
represented as the last species $K+2$ in our model. We are unaware of other
all-spheres models that deal explicitly with the voids between spheres. We
suspect that including such voids is needed because voids are in different
amounts depending on the composition of the solution and can move in any
system of spheres crowded into a small space.

Obviously, all properties of water cannot be represented this way: water is a
highly charged molecule although its net charge is zero, and polymeric
structures can exist and may be important, along with hydrogen bonds of the
low or high energy type \cite{SW04,DS01}. Moreover, not all defects in packing
can be represented by a single void site, just as not all properties of water
can be represented by uncharged spheres. The question is whether a model that
includes only the excluded volume of water and a continuous void space between
particles is able to deal with the selectivity data of the calcium channel. We
will see that it can.

The total volume $V$ of the system consists of the volumes of all particles
and the total void volume $v_{K+2}$, i.e., $V=\sum_{j=1}^{K+1}v_{j}%
N_{j}+v_{K+2}$. Under the bulk condition, dividing this equation by $V$ yields
the bulk void volume fraction
\begin{equation}
\Gamma^{\text{B}}=\frac{v_{K+2}}{V}=1-\sum_{j=1}^{K+1}v_{j}\frac{N_{j}}%
{V}=1-\sum_{j=1}^{K+1}v_{j}C_{j}^{\text{B}} \tag{3}%
\end{equation}
expressed in terms of the nonuniform particle volumes $v_{j}$ and the bulk
concentrations $C_{j}^{\text{B}}$ of all particle species. We are aware that a
model of this sort can be extended into an all-spheres model of ionic
solutions of the so called bio-ions Na$^{+}$, K$^{+}$, Ca$^{2+}$, and Cl$^{-}
$.

Using the Stirling formula $\ln M!\approx M\ln M-M$ with $M>>1$, the
electrochemical potential of particle species $i=1,\cdots,K+1$ is%
\begin{equation}
\mu_{i}=\frac{\partial F(N)}{\partial N_{i}}=q_{i}\phi+k_{B}T\ln\frac{N_{i}%
}{N-\sum_{j=1}^{K+1}N_{j}} \tag{4}%
\end{equation}
from which we deduce global probabilities $P_{i}=N_{i}/N$ for all particle
species. If we extend our theory by introducing local probabilities
$p_{i}(\mathbf{r})=v_{i}C_{i}(\mathbf{r})$ that depend on location, in effect
allowing probabilities to depend on location as in the theory of stochastic
processes \cite{S09} (applied for example to ionic channels
\cite{EK95,SN01,SS04}), the electrochemical potential can be generalized
locally to
\begin{align}
\mu_{i}(\mathbf{r})  &  =q_{i}\phi(\mathbf{r})+k_{B}T\ln\frac{v_{i}%
C_{i}(\mathbf{r})}{1-\sum_{j=1}^{K+1}v_{j}C_{j}(\mathbf{r})}=q_{i}%
\phi(\mathbf{r})+k_{B}T\ln\frac{C_{i}(\mathbf{r})}{C_{i}^{\text{B}}}+\mu
_{i}^{\text{ex}}(\mathbf{r})\tag{5}\\
\mu_{i}^{\text{ex}}(\mathbf{r})  &  =k_{B}T\ln\frac{v_{i}C_{i}^{\text{B}}%
}{\Gamma(\mathbf{r)}}\text{, \ \ \ }\Gamma(\mathbf{r)}=1-\sum_{j=1}^{K+1}%
v_{j}C_{j}(\mathbf{r})=v_{K+2}C_{K+2}(\mathbf{r})\text{,} \tag{6}%
\end{align}
where $C_{i}(\mathbf{r})$ is the concentration function of spatial variable
$\mathbf{r}$ in the solvent domain $\Omega_{s}$, $\mu_{i}^{\text{ex}%
}(\mathbf{r})$ is the excess chemical potential, and $\Gamma(\mathbf{r)}$ is
the void fraction function with $C_{K+2}(\mathbf{r})$ representing the
distribution function of interstitial voids. When $\phi=0$, $C_{i}%
(\mathbf{r})=C_{i}^{\text{B}}$ and hence $\mu_{i}^{\text{ex}}=\mu
_{i}^{\text{B}}=k_{B}T\ln\left(  v_{i}C_{i}^{\text{B}}/\Gamma^{\text{B}%
}\right)  $ is a constant.

\textbf{\underline{\textbf{The excess chemical potential}} }is a measure of
nonideality that helps understand qualitative behavior. For example, the
larger the size $v_{i}$ of a type $i$ particle, the larger is the activation
barrier $\mu_{i}^{\text{ex}}(\mathbf{r})$ and the harder it is for the
particle to make a transition at $\mathbf{r}$ from a local minimum of $\mu
_{i}^{\text{ex}}$ to another local minimum nearby \cite{B13}. The transition
mechanism is related to the vacancy configuration as well, i.e., the smaller
the value of $\Gamma(\mathbf{r)}$, the more crowded the ions are at
$\mathbf{r}$, the harder transition. The excess chemical potential is closely
related to the sizes of all particles $v_{j}$, their interstitial voids
$\Gamma(\mathbf{r)}$, their configurations $C_{j}(\mathbf{r})$, as well as
their bulk concentrations $C_{j}^{\text{B}}$.

To our knowledge, all existing continuum models do not explicitly take into
account the finite size effect of water, let alone the effect of interstitial
voids. We did not consider these two effects in our previous work \cite{LE13}
in which water was treated as a single continuous dielectric medium without
any voids and the resulting electrochemical potential $\mu_{i}(\mathbf{r})$
was shown to be a mathematical description of the primitive model of
electrolytes, as used in most Monte Carlo and density functional theory
models. Our continuum primitive model could well match Monte Carlo (discrete
primitive model) results that were obtained in equilibrium state. However, as
we proceeded to study nonequilibrium systems using this primitive model, we
had difficulty computing the experimental currents reported in \cite{AM84} due
to either inconsistent physics or divergent numerics.

\textbf{\underline{\textbf{The calcium channel}}} operates very delicately in
physiological and experimental conditions as it shifts its exquisitely tuned
conductance from Na$^{+}$-flow, to Na$^{+}$-blockage, and to Ca$^{2+}$-flow
when bath Ca$^{2+}$ concentration varies from $10^{-10}$ to $10^{-2}$ M. The
$10^{8}$-fold range of experimental conditions make modeling and numerical
implementation very challenging. This huge dynamic range was accommodated in
our previous work by using an artificial potential to confine mobile oxygen
ions of side chains within a filter region, just as that used in all Monte
Carlo simulations on the same channel (see e.g. \cite{MG09}). The artificial
potential hindered our effort to match the experimental data since it is a
gross approximation of the constraining energy needed to keep the protein
atoms in the filter region without specifically considering the void effect.
Indeed, using a restraining potential can lead to inconsistencies, since
maintaining the steric and electrical potential as conditions change requires
injection of energy and charge into the system \cite{E96}. We obtain
convergent and consistent results using the steric potential in place of the
artificial constraining potential of earlier models. The steric potential is
an output of our model and varies automatically as conditions change. It has
the same units as a confining potential but is as different as the voltages at
an input and an output of an ideal amplifier. The following analysis shows
that the void species in our model is important not only to describe a
consistent physics of the steric potential but also to compute the steric
energy that can reflect the $10^{8}$-fold experimental conditions. It will be
interesting to examine the properties of a model of bulk ionic solutions that
contains voids calculated consistently in an all-spheres model of ions and water.

Setting $\mu_{i}(\mathbf{r})=\mu_{i}^{\text{B}}$ (see below for physical
reason), the concentration of species $i$ particles can be expressed by the
Fermi like distribution function%
\begin{equation}
C_{i}(\mathbf{r})=C_{i}^{\text{B}}\exp\left(  -\beta_{i}\phi(\mathbf{r}%
)+S^{\text{trc}}(\mathbf{r})\right)  \text{, \ \ }S^{\text{trc}}%
(\mathbf{r})=\ln\frac{\Gamma(\mathbf{r)}}{\Gamma^{\text{B}}}, \tag{7}%
\end{equation}
where $\beta_{i}=q_{i}/k_{B}T$ and $S^{\text{trc}}(\mathbf{r})$ is called the
steric potential that describes the combined effect of all excess chemical
potentials $\mu_{j}^{\text{ex}}$ of all particle species $j=1,\cdots K+1$. The
distribution (7) is of Fermi type since all concentration functions
\begin{equation}
C_{i}(\mathbf{r})=\frac{\alpha_{i}-\alpha_{i}\sum_{j\neq i}^{K+1}v_{j}%
C_{j}(\mathbf{r})}{1+\alpha_{i}v_{i}}<\frac{1}{(1/\alpha_{i})+v_{i}}=\frac
{1}{v_{i}}\text{,} \tag{8}%
\end{equation}
$i=1,\cdots,K+1$, are bounded from above with $\alpha_{i}=C_{i}^{\text{B}}%
\exp\left(  -\beta_{i}\phi(\mathbf{r})\right)  /\Gamma^{\text{B}}>0$, i.e.,
$C_{i}(\mathbf{r})$ cannot exceed the maximum value $1/v_{i}$ for any
arbitrary (or even infinite) potential $\phi(\mathbf{r})$ at any location
$\mathbf{r}$ in the domain $\Omega_{s}$. In this mean-field Fermi
distribution, it is impossible for a volume $v_{i}$ to be completely filled
with particles, i.e., it is impossible to have $v_{i}C_{i}(\mathbf{r})=1$ (and
thus $\Gamma(\mathbf{r)}=0$) since that would make the excess chemical
potential $\mu_{i}^{\text{ex}}$ infinitely large or $S^{\text{trc}}=-\infty$
and hence $C_{i}(\mathbf{r})=0$, a contradiction. For this reason, \textbf{we
must include the voids as a separate species if water is treated as hard
spheres}. Otherwise, the volume $v_{i}$ would be easily filled by particles in
the mean-field sense at moderate electric potential such that the steric
potential would be unphysical. The requirement of voids when all particles are
represented as hard spheres will be justified again from a viewpoint of Gibbs'
free energy.

The classical Boltzmann distribution appears if all particles are treated as
volumeless points, i.e., $v_{i}=0$ and $\Gamma(\mathbf{r})=\Gamma^{\text{B}%
}=1$. It may produce an infinite concentration $C_{i}(\mathbf{r}%
)\rightarrow\infty$ in crowded conditions when $-\beta_{i}\phi(\mathbf{r}%
)\rightarrow\infty$, close to charged surfaces for example, an impossible
result \cite{LE13,L13,LE14}. The difficulty in the application of classical
Boltzmann distributions to saturating systems has been avoided in the
physiological literature (apparently starting with Hodgkin, Huxley, and Katz
\cite{HH49}) by redefining the Boltzmann distribution to deal with systems
that can only exist in two states. This redefinition has been vital to
physiological research and is used in hundreds of papers \cite{B00,BV13}, but
confusion results when the physiologists' saturating two-state Boltzmann is
not kept distinct from the unsaturating Boltzmann distribution of statistical
mechanics \cite{M76}.

To further account for the correlation effect of ions and the screening effect
of water molecules, we have developed efficient 3D methods \cite{L13} for
solving the Poisson-Fermi (PF) equation \cite{LE13,L13,LE14,S06,BS11}
\begin{equation}
\epsilon_{s}\left(  l_{c}^{2}\nabla^{2}-1\right)  \nabla^{2}\phi
(\mathbf{r})=\sum_{i=1}^{K}q_{i}C_{i}(\mathbf{r})=\rho(\mathbf{r}) \tag{9}%
\end{equation}
self-consistently with Eq. (7) for $\phi(\mathbf{r})$, where $l_{c}$ is a
correlation length \cite{S06,BS11}, $\epsilon_{s}=\epsilon_{\text{w}}%
\epsilon_{0}$, $\epsilon_{\text{w}}$ is a dielectric constant of water in the
bath, and $\epsilon_{0}$ is the vacuum permittivity. The fourth-order PF
equation reduces to the classical Poisson-Boltzmann (PB) equation when
$l_{c}=S^{\text{trc}}(\mathbf{r})=0$. If $l_{c}\neq0$, the dielectric operator
$\widehat{\epsilon}=\epsilon_{s}(1-l_{c}^{2}\nabla^{2})$ is used to
approximate the permittivity of the bulk solvent and the linear response of
correlated ions \cite{BS11}. The dielectric function $\widetilde{\epsilon
}(\mathbf{r})=\epsilon_{s}/(1+\eta/\rho)$ is a further approximation of
$\widehat{\epsilon}$. It is found by transforming Eq. (9) into two
second-order equations $\epsilon_{s}\left(  l_{c}^{2}\nabla^{2}-1\right)
\Psi=\rho$ and $\nabla^{2}\phi=\Psi$. We introduce a density like variable
$\Psi$ that yields a polarization charge density $\eta=-\epsilon_{s}\Psi-\rho$
using Maxwell's first equation \cite{LE13,L13}.

The free energy formula (1) is useful for a thermodynamic system that involves
a limited number of particles for MD or MC simulations particularly without
flow, or spatially nonuniform boundary conditions. If the system is
nonequilibrium or has numerous particles and complicated boundary conditions,
the PF equation (9) will be more suitable for theoretical investigation.

\textbf{\underline{\textbf{Free energy functional.}}} We look at our model now
from the perspective of a generalization of free energy that we call the
Gibbs-Fermi free energy. The PF equation is a minimizer of the following Gibbs
free energy functional%
\begin{align}
G^{\text{Fermi}}  &  =\int_{\Omega_{s}}d\mathbf{r}\left\{  -\frac{\epsilon
_{s}l_{c}^{2}}{2}\left(  \nabla^{2}\phi\right)  ^{2}-\frac{\epsilon_{s}}%
{2}\left\vert \nabla\phi\right\vert ^{2}+\rho\phi+g\right\} \tag{10}\\
g  &  =k_{B}T\left(  \sum_{j=1}^{K+1}\left[  C_{j}\ln\left(  v_{j}%
C_{j}\right)  -C_{j}-C_{j}\ln\left(  v_{K+2}C_{K+2}\right)  +\frac{\lambda
_{j}C_{j}}{k_{B}T}\right]  \right) \nonumber
\end{align}
by taking energy variations with respect to $\phi$, i.e., $\frac{\delta
G^{\text{Fermi}}}{\delta\phi}=0$. The Fermi distribution (7) follows from
$\frac{\delta G^{\text{Fermi}}}{\delta C_{i}}=0$, where
\begin{equation}
\lambda_{i}=-\mu_{i}^{\text{B}}=-k_{B}T\ln\frac{v_{i}C_{i}^{\text{B}}}%
{\Gamma^{\text{B}}} \tag{11}%
\end{equation}
is the Lagrange multiplier for the mass conservation (the total number $N_{i}
$) of particle species $i$ \cite{WZ12}. The minimization $\frac{\delta
G^{\text{Fermi}}}{\delta C_{i}}=0$ is equivalent to setting $\mu
_{i}(\mathbf{r})=\mu_{i}^{\text{B}}$ for Eq. (7) with the identity
$\Gamma=v_{K+2}C_{K+2}$. The electrochemical potential (5) can also be defined
by the functional as
\begin{equation}
\mu_{i}=\frac{\delta G^{\text{Fermi}}}{\delta C_{i}}+\mu_{i}^{\text{B}}.
\tag{12}%
\end{equation}

When $l_{c}=S^{\text{trc}}(\mathbf{r})=v_{j}=0$, $j=1,\cdots,K+1$ (without
correlation and steric terms), this functional yields the PB equation
$-\epsilon_{s}\nabla^{2}\phi(\mathbf{r})=\rho(\mathbf{r})$ and the Boltzmann
distribution $C_{i}(\mathbf{r})=C_{i}^{\text{B}}\exp\left(  -\beta_{i}%
\phi(\mathbf{r})\right)  $ since $v_{K+2}C_{K+2}=\Gamma^{\text{B}}=1$ and
$g=k_{B}T\sum_{j=1}^{K+1}\left[  C_{j}\ln(C_{j}/C_{j}^{\text{B}}%
)-C_{j}\right]  $. Note that all $v_{j}$ in $g$ are canceled before setting
$v_{j}=0$ since%
\begin{equation}
C_{j}\ln(v_{j}C_{j})+\lambda_{j}\frac{C_{j}}{k_{B}T}=C_{j}\ln(v_{j}%
C_{j})-C_{j}\ln\frac{v_{j}C_{j}^{\text{B}}}{\Gamma^{\text{B}}}=C_{j}\ln
\frac{C_{j}\Gamma^{\text{B}}}{C_{j}^{\text{B}}}. \tag{13}%
\end{equation}
We need $v_{j}$ in $g$ to justify that the local electrochemical potential (5)
can be defined by the functional (10) and that the Fermi distribution (7) is a
consequence of mass conservation by (12).

\textbf{\underline{\textbf{Voids are needed.}}} To establish a consistent
generalization from Boltzmann to Fermi distribution it is critical to express
the energy functional (10) by means of the void fraction function
$v_{K+2}C_{K+2}(\mathbf{r})=1-\sum_{j=1}^{K+1}v_{j}C_{j}(\mathbf{r})$ in $g$.
Otherwise, $\ln\left(  v_{K+2}C_{K+2}(\mathbf{r})\right)  =\ln1=0$ (without
the void term) implies that $\sum_{j=1}^{K+1}v_{j}C_{j}(\mathbf{r})=0$, i.e.,
all $v_{j}=0$ and the Boltzmann functional of volumeless particles that we are
seeking to replace. This means that \textbf{it is impossible to treat all ions
and water molecules as hard spheres and at the same time achieve a zero volume
of interstitial voids between all particles}. Therefore, the Gibbs-Fermi
functional (10) is not only consistent but also needed (and of course
physical) with either $v_{j}=0$ (all particles are volumeless points) or
$v_{j}\neq0$ (all particles are spheres).

Of course, we could treat ions as spheres but water as a continuous medium
(without voids) that then forms the single site in Eq. (2) in place of the
voids as previously proposed in our paper \cite{LE13} for the primitive
solvent model. The void fraction $\Gamma(\mathbf{r})$ would then become the
water fraction \ $\Gamma(\mathbf{r)}=1-\sum_{j=1}^{K}v_{j}C_{j}(\mathbf{r})$,
where the upper limit is $K$ instead of $K+1$. This is precisely the primitive
model implemented in the Monte Carlo simulations of Boda and Henderson.
Consequently, the primitive model may yield incorrect water densities,
pressures, and dielectric coefficient in mean-field sense for nonideal and
inhomogeneous systems (see Section 5 for more details). This important
limitation in the continuous water version of the all-spheres model was
pointed out early in its history \cite{NG01}.

\textbf{\underline{\textbf{Comparison with other treatments of finite sized
particles.}}} All existing free energy functionals that specifically include
either uniform size ($v=v_{j}$ for all $j$) \cite{BA97,L09,SB10,BS11,LZ11} or
nonuniform sizes ($v_{i}\neq v_{j}$) \cite{ZW11,L13,QT14} cannot reduce to
their corresponding Boltzmann functionals by directly setting $v_{j}=0$
\cite{L09,QT14} because those functionals retain the local probability form of
$v_{j}C_{j}\ln\left(  v_{j}C_{j}\right)  =p_{j}\ln p_{j}$ in their Gibbs
entropy. They use an inconsistent reciprocal term involving a \textit{uniform}
particle size, namely $1/v$, instead of a consistent term involving the
nonuniform particle sizes. The local probability $p_{j}(\mathbf{r})$ of any
particle species $j$ in our Gibbs entropy
\begin{equation}
-k_{B}\sum_{j=1}^{K+1}C_{j}\ln\left(  v_{j}C_{j}\right)  -C_{j}\ln\left(
v_{K+2}C_{K+2}\right)  +\frac{\lambda_{j}C_{j}}{k_{B}T}=-k_{B}\sum_{j=1}%
^{K+1}C_{j}\ln\frac{C_{j}\Gamma^{\text{B}}}{C_{j}^{\text{B}}\Gamma} \tag{14}%
\end{equation}
is instead expressed in terms of the local probability ratio $C_{j}%
(\mathbf{r})/\Gamma(\mathbf{r})$ and the global probability ratio
$\Gamma^{\text{B}}/C_{j}^{\text{B}}$ between the particle fraction
(probability) $C_{j}(\mathbf{r})$ and the void fraction $\Gamma(\mathbf{r})$
per unit volume. In other words, the local probability $p_{j}(\mathbf{r})$ in
the Gibbs-Fermi treatment changes with varying configurations of all particles
($\Gamma(\mathbf{r})=1-\sum_{j=1}^{K+1}v_{j}C_{j}(\mathbf{r})$) and voids
($\Gamma(\mathbf{r})=v_{K+2}C_{K+2}(\mathbf{r})$). The local probability at
any location, including the binding site, is also connected to the bulk
conditions in the bath as implied by Eq. (7). It also depends implicitly on
the sizes of all particles, valences of ionic species, and long range
(Coulomb) as well as short range (Lennard-Jones) distances between all
particles. All these physical properties are lumped into the steric potential
functional $S^{\text{trc}}(\mathbf{r})=\ln\frac{\Gamma(\mathbf{r)}}%
{\Gamma^{\text{B}}}$ in a very simple and unified way.

The void fractions $\Gamma^{\text{B}}$ and $\Gamma$ represent the
Lennard-Jones distances between all particles in a mean-field approximation.
More specifically, the L-J potential $V(r)=4\left(  (\sigma/r)^{12}%
-(\sigma/r)^{6}\right)  $ \cite{V67} can be used to determine the distance $r$
between any pair of particles. In bulk solutions, the distance $r=\sigma$
yields $V(r)=0$ that corresponds to a finite but fixed distance $\sigma$
between any two adjacent particles in the system and thus to the constant bulk
void fraction $\Gamma^{\text{B}}$. Similarly, the nonuniform void function
$\Gamma(\mathbf{r})$ corresponds to nonuniform inter-particle distances $r$
that may or may not equal $\sigma$ for all pairs of adjacent particles.
Nonzero L-J potentials are in general highly oscillatory and extremely
expensive to compute in a system of numerous particles. Including external
fields adds problems of consistency with spatially nonuniform far field
boundary conditions to the problems of computational expense. The void
function $\Gamma(\mathbf{r})$ or equivalently the steric potential
$S^{\text{trc}}(\mathbf{r})$ is on the other hand quite smooth and relatively
very easy to compute. The convolutional density functional on any pair of
concentration functions $C_{i}(\mathbf{r})$ and $C_{j}(\mathbf{r}^{\prime})$
with a L-J kernel or DFT representation of the interaction potential proposed
by Eisenberg et al. \cite{EH10} is another mean-field approximation, which is
more accurate (but much more difficult to compute reliably) than the steric
potential $S^{\text{trc}}(\mathbf{r})$ since the convolutional functional is
nonlocal whereas $S^{\text{trc}}(\mathbf{r})$ is local. The local steric
potential of this paper can be used in place of the nonlocal L-J or DFT
potential of \cite{EH10} and therefore the energy variational theory based on
the Onsager dissipation principle developed in \cite{EH10} can be applied to
the Gibbs-Fermi functional (10).

\section{Poisson-Nernst-Planck-Fermi Theory}

For nonequilibrium systems, the classical PNP model \cite{CB92,EC93,EK95} can
then be generalized to the Poisson-Nernst-Planck-Fermi model by coupling the
flux density equation
\begin{equation}
\frac{\partial C_{i}(\mathbf{r},t)}{\partial t}=-\nabla\cdot\mathbf{J}%
_{i}(\mathbf{r},t),\text{ }\mathbf{r}\in\Omega_{s} \tag{15}%
\end{equation}
of each particle species $i=1,\cdots,K+1$ (including water) to the PF equation
(9), where the flux density is defined as%
\begin{equation}
\mathbf{J}_{i}(\mathbf{r},t)=-D_{i}\left[  \nabla C_{i}(\mathbf{r}%
,t)+\beta_{i}C_{i}(\mathbf{r},t)\nabla\phi(\mathbf{r},t)-C_{i}(\mathbf{r}%
,t)\nabla S^{\text{trc}}(\mathbf{r},t)\right]  \text{,} \tag{16}%
\end{equation}
$D_{i}$ is the diffusion coefficient, and the time variable $t$ is added to
describe the dynamics of electric $\phi(\mathbf{r},t)$ and steric
$S^{\text{trc}}(\mathbf{r},t)$ potentials. The flux equation (15) is called
the Nernst-Planck-Fermi equation because the Fermi steric potential
$S^{\text{trc}}(\mathbf{r},t)$ is introduced to the classical NP equation.

At equilibrium, the net flow of each particle species is a zero vector, i.e.,
$\mathbf{J}_{i}(\mathbf{r})=\mathbf{0}$ (in a steady state) which implies that
$C_{i}\exp(\beta_{i}\phi-S^{\text{trc}})=$ const. $=C_{i}^{\text{B}}$ for
$\phi=S^{\text{trc}}=0$. Therefore, the NPF equation (15) reduces to the Fermi
distribution (7) at equilibrium. Similarly, the classical NP equation reduces
to the Boltzmann distribution at equilibrium.

\textbf{\underline{\textbf{The steric force.}}} The gradient of the steric
potential $\nabla S^{\text{trc}}$ in (16) represents an entropic force of
vacancies exerted on particles. The negative sign in $-C_{i}\nabla
S^{\text{trc}}$ means that the steric force $\nabla S^{\text{trc}}$ is in the
opposite direction to the `diffusional' force $\nabla C_{i}$.

The larger $S^{\text{trc}}(\mathbf{r})=\ln\frac{\Gamma(\mathbf{r)}}%
{\Gamma^{\text{B}}}$ (meaning more space available to the particle as implied
by the numerator) at $\mathbf{r}$ in comparison with that of neighboring
locations, the more the entropic force pushes the particle to the location
$\mathbf{r}$, which is simply the opposite mechanism of the diffusional force
$\nabla C_{i}(\mathbf{r})$ that pushes the particle away from $\mathbf{r}$ if
the concentration is larger at $\mathbf{r}$ than that of neighboring
locations. Moreover, the Nernst-Einstein relationship \cite{H01} implies that
the steric flux $D_{i}C_{i}\nabla S^{\text{trc}}$ is greater if the particle
is more mobile. Therefore, the gradients of electric and steric potentials
($\nabla\phi$ and $\nabla S^{\text{trc}}$) describe the charge/space
competition mechanism of particles in a crowded region within a mean-field
framework. Since $S^{\text{trc}}(\mathbf{r},t)$ describes the dynamics of void
movements, the dynamic crowdedness (pressure) of the flow system can also be quantified.

The motion of water molecules is directly controlled by the steric potential
in our model and their distributions are expressed by $C_{K+1}(\mathbf{r}%
,t)=C_{K+1}^{\text{B}}\exp\left(  S^{\text{trc}}(\mathbf{r},t)\right)  $.
Nevertheless, this motion is still implicitly affected by the electric
potential $\phi(\mathbf{r},t)$ via the correlated motion of ions described by
other $C_{j}(\mathbf{r},t)$ in the void fraction function $\Gamma
(\mathbf{r},t)$ and hence in the charge density $\rho$ in (9).

\textbf{\underline{Water is polarizable in our model.}} From (9), $\nabla
^{2}\phi=\Psi$, and $\eta=-\epsilon_{s}\Psi-\rho$, we deduce that the Poisson
equation%
\begin{equation}
-\epsilon_{s}\nabla^{2}\phi=\rho-\epsilon_{s}l_{c}^{2}\nabla^{2}\Psi=\rho
+\eta\text{,} \tag{17}%
\end{equation}
which describes the electric field $\mathbf{E}=-\nabla\phi$ in the system,
contains the charge source not only from the ions ($\rho=\sum_{i=1}^{K}%
q_{i}C_{i}$) but also from the polar water ($\eta$) provided that the
correlation length $l_{c}\approx l_{B}q_{i}^{2}$ is not zero. The polarization
charge density $\eta$ is proportional to the fourth order of the ionic valence
$z_{i}$. The fourth order dependence shows that even in a mean field theory
valency of ions is expected to play an important role as known in chemical and
biological systems \cite{BF02,SM03,BV09,BS11,L13,LE13,LE14}.

In summary, the PNPF model accounts for the steric effect of ions and water
molecules, the correlation effect of crowded ions, the screening effect of
polar water, as well as the charge/space competition effect of ions and water
molecules of different sizes and valences. These effects are all closely
related to the interstitial voids between particles and described by two
additional terms, namely, the correlation length and the steric potential.

\section{A Molecular-Continuum Model of a Calcium Channel}

To test the PNPF theory, we use the Lipkind-Fozzard molecular model
\cite{LF01} of L-type Ca channels in which the EEEE locus (four glutamate side
chains modeled by 8 O$^{1/2-}$ ions) forms a high-affinity Ca$^{2+}$ binding
site that is essential to Ca$^{2+}$ selectivity, blockage, and permeation. We
refer to Fig. 9 in \cite{LF01} or Fig. 1 in \cite{LE14} for a 3D illustration
of the EEEE locus. A 2D cross section of a simplified 3D channel geometry for
the present work is shown in Fig. 1, where the central circle denotes the
binding site, the other four circles denote the side view of 8 O$^{1/2-}$
ions, $\Omega_{s}$ is the solvent region consisting of two baths and the
channel pore including the binding domain $\overline{\Omega}_{\text{Bind}}$,
$\partial\Omega_{s}$ is the solvent boundary, and $\partial\Omega
_{\text{Bath}}$ is the outside and inside bath boundary. Fig. 2 is a sketch of
the binding site and O$^{1/2-}$ ions, where $d_{O}^{Ca}$ is the distance
between the center of a binding Ca$^{2+}$ ion and the center $c_{j}$ of any
O$^{1/2-}$, and $A$ is any point on the surface of the site. In our model, the
8 O$^{1/2-}$ ions are not contained in the solvent region $\Omega_{s}$.
Particle species are indexed by 1, 2, 3, and 4 for Na$^{+}$, Ca$^{2+}$,
Cl$^{-}$, and H$_{2}$O with radii $a_{1}=a_{\text{Na}^{+}}=$ $0.95$,
$a_{2}=a_{\text{Ca}^{2+}}=0.99$, $a_{3}=r_{\text{Cl}^{-}}=1.81$, and
$a_{4}=a_{\text{H}_{2}\text{O}}=1.4$ \AA , respectively.%
\begin{figure}
[ptb]
\begin{center}
\includegraphics[
height=2.7077in,
width=4.6164in
]%
{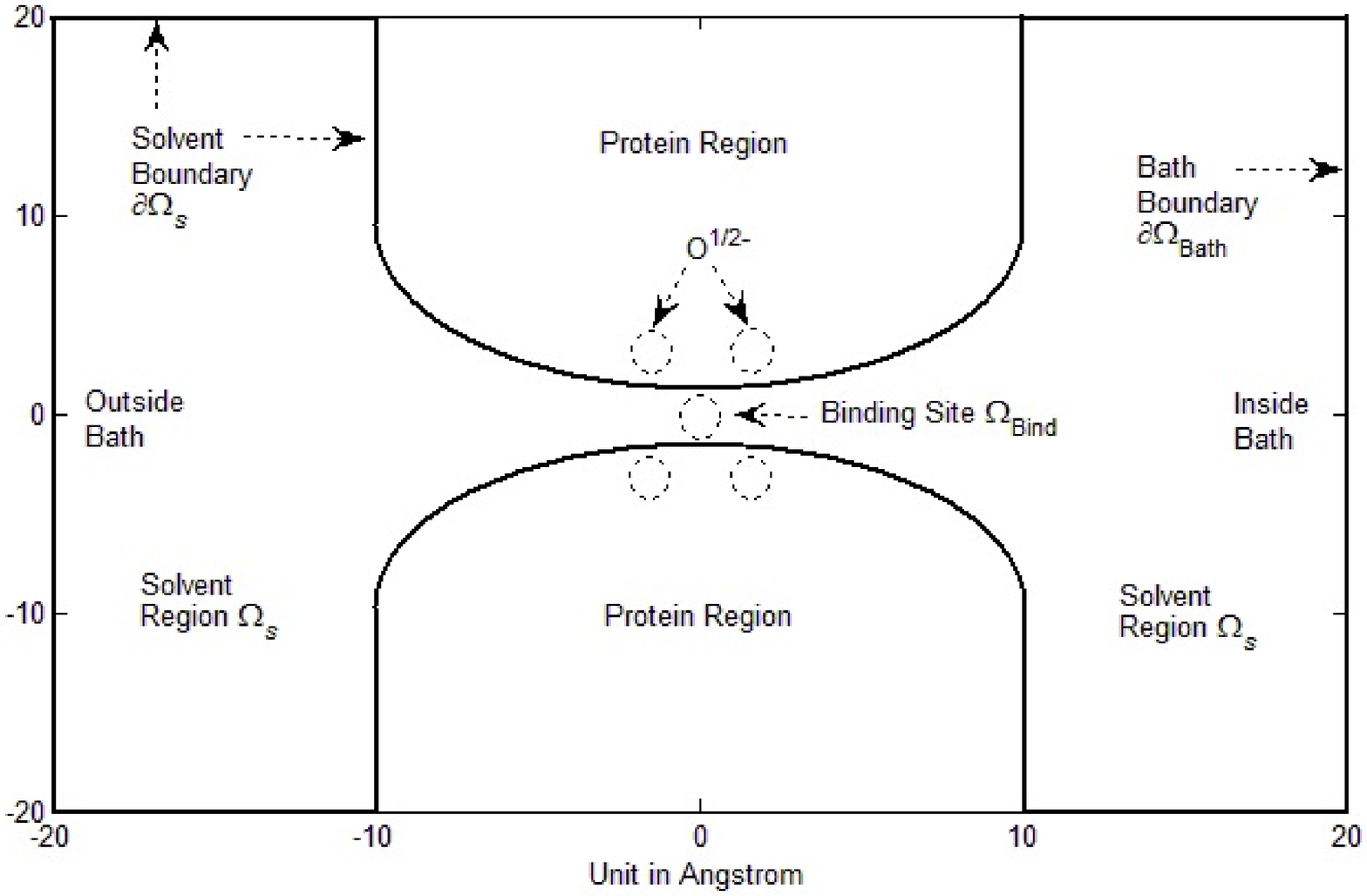}%
\caption{A simplified Ca channel geometry with baths, pore, and binding site.
The channel is placed in a cubic box with the length of each side being 40
\AA . The solvent region $\Omega_{s}$ consists of two baths and the channel
pore with the boundary $\partial\Omega_{s}$. The binding site $\overline
{\Omega}_{\text{Bind}}$ is contained in $\Omega_{s}$ but the O$^{1/2-}$ ions
are not in $\Omega_{s}$. The outside and inside bath boundary is denoted by
$\partial\Omega_{\text{Bath}}$.}%
\end{center}
\end{figure}
\begin{figure}
[ptbptb]
\begin{center}
\includegraphics[
height=2.7285in,
width=4.651in
]%
{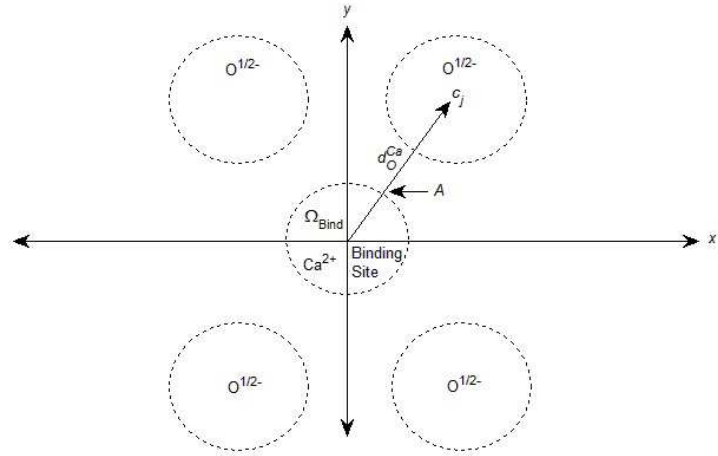}%
\caption{The binding distance between the center of the binding Ca$^{2+}$ ion
and the center $c_{j}$ of the $j^{\text{th}}$ O$^{1/2-}$ ion is denoted by
$d_{O}^{Ca}$ for $j=1,\cdots,8$. $A$ is any point on the surface of the
binding ion.}%
\end{center}
\end{figure}

In \cite{LE14}, we proposed an algebraic model for calculating the electrical
potential $\overline{\phi}_{b}$ and the steric potential $\overline{S}%
_{b}^{\text{trc}}$ in $\overline{\Omega}_{\text{Bind}}$ by using Coulomb's law
with the atomic structure of binding ion and atoms in a channel protein as
shown in Fig. 2, without solving the Poisson-Fermi equation (9) in
$\overline{\Omega}_{\text{Bind}}$. The binding potential $\overline{\phi}_{b}$
is then used as a Dirichlet boundary condition in $\overline{\Omega
}_{\text{Bind}}$ for solving the PF equation in the solvent region between the
bath and binding boundary, i.e., in $\Omega_{s}\backslash\overline{\Omega
}_{\text{Bind}}$, to obtain the potential profile $\phi(\mathbf{r})$ that
connects $\overline{\phi}_{b}$ in $\overline{\Omega}_{\text{Bind}}$ to the
potential $V_{\text{i}}$ (or $V_{\text{o}}$) on the inside (or outside) bath boundary.

The filter domain defined in \cite{LE14} is simply taken to be the binding
domain $\overline{\Omega}_{\text{Bind}}$ in this paper. The volume of this
domain is an unknown variable $v_{b}$ that changes with different charges in
the binding site. We do not define an ad hoc filter for which its volume is
fixed to one value in one implementation and possibly to another value in
other implementation. We show that the variable binding volume $v_{b}$ plays
an essential role in determining the steric potential in and around the
binding site and consequently the hydrophobicity of the EEEE locus under
different bath conditions.

The algebraic model \cite{LE14} is defined in $\overline{\Omega}_{\text{Bind}%
}$ and consists of the following equations
\begin{equation}
\left\{
\begin{array}
[c]{l}%
O_{1}^{b}=v_{b}C_{1}^{\text{B}}\exp(-\beta_{1}\overline{\phi}_{b}+\overline
{S}_{b}^{\text{trc}})\\
O_{2}^{b}=v_{b}C_{2}^{\text{B}}\exp(-\beta_{2}\overline{\phi}_{b}+\overline
{S}_{b}^{\text{trc}})\\
O_{4}^{b}=v_{b}C_{4}^{\text{B}}\exp(\overline{S}_{b}^{\text{trc}})
\end{array}
\right.  , \tag{18}%
\end{equation}%
\begin{equation}
\overline{S}_{b}^{\text{trc}}=\ln\frac{v_{b}-v_{1}O_{1}^{b}-v_{2}O_{2}%
^{b}-v_{4}O_{4}^{b}}{v_{b}\Gamma^{\text{B}}} \tag{19}%
\end{equation}%
\begin{equation}
\frac{e}{4\pi\epsilon_{0}}\left(  \sum_{j=1}^{8}\frac{z_{\text{O}^{1/2-}}%
}{|c_{j}-A|}+\frac{O_{1}^{b}z_{\text{Na}^{+}}}{a_{\text{Na}^{+}}}+\frac
{O_{2}^{b}z_{\text{Ca}^{2+}}}{a_{\text{Ca}^{2+}}}\right)  =\overline{\phi}%
_{b}, \tag{20}%
\end{equation}
where $O_{1}^{b}$, $O_{2}^{b}$, and $O_{4}^{b}$ denote the occupancy numbers
of Na$^{+}$, Ca$^{2+}$, and H$_{2}$O in $v_{b}$, respectively, $\overline
{\phi}_{b}$ and $\overline{S}_{b}^{\text{trc}}$ are average electrical and
steric potentials, and $|c_{j}-A|$ is the distance between $A$ and $c_{j}$ in
Fig. 2.

In this mean field, we allow $O_{1}^{b}$ and $O_{2}^{b}$ (and hence the total
charge $O_{1}^{b}ez_{\text{Na}^{+}}+O_{2}^{b}ez_{\text{Ca}^{2+}}$) to vary
continuously subject to the condition on their sum $O_{1}^{b}+O_{2}^{b}=1$ in
the binding volume $v_{b}$. Eqs. (18) and (19) uniquely determine the four
unknowns $v_{b}$, $O_{4}^{b}$, $\overline{\phi}_{b}$ and $\overline{S}%
_{b}^{\text{trc}}$ with $O_{1}^{b}$ and $O_{2}^{b}$ being given. Eq. (20)
uniquely determines the locations ($c_{j}$) of 8 O$^{1/2-}$ ions (and thus the
binding distance $d_{O}^{Ca}$ or $d_{O}^{Na}$ in Fig. 2) once $\overline{\phi
}_{b}$ is obtained. Note that the binding distance $d_{O}^{O_{1}^{b}%
Na+O_{2}^{b}Ca}$ (or $c_{j}$) changes continuously with varying $O_{1}^{b}$
and $O_{2}^{b}$ but $\overline{\phi}_{b}$ remains fixed, where the binding ion
$O_{1}^{b}Na+O_{2}^{b}Ca$ is a linear combination of Na$^{+}$ and Ca$^{2+}$.
Therefore, O$^{1/2-}$ ions are movable --- the protein is flexible in our
model --- as their locations $c_{j}$ changes with varying $O_{1}^{b}$ and
$O_{2}^{b}$ \cite{LE14}. Note that we change the probability notation $P_{i}$
in \cite{LE14} to the occupancy notation $O_{i}^{b}$ to reflect the
deterministic, instead of probabilistic, nature of the PNPF continuum model.
In this simple algebraic model, we do not consider the hydrogen ions that may
react with carboxyl anions in the protein. Experiments done at pH 8 (as many
have been done) do not involve association of hydrogen ions with carboxyl anions.

For the half-blockage experimental condition \cite{AM84}%
\begin{equation}
\underset{\text{{\normalsize Experimental Data}}}{\underbrace{C_{\text{Na}%
^{+}}^{\text{B}}=C_{\text{1}}^{\text{B}}=32\text{ mM, }C_{\text{Ca}^{2+}%
}^{\text{B}}=C_{\text{2}}^{\text{B}}=0.9\text{ }\mu\text{M,}}} \tag{21}%
\end{equation}
we follow convention and assume relative occupancies of a filled channel,
$O_{1}^{b}=0.5$ and $O_{2}^{b}=0.5$, and thereby obtain $\overline{\phi}%
_{b}=-10.48$ $k_{B}T/e$, $\overline{S}_{b}^{\text{trc}}=-1.83$, and
$v_{b}=4.56$ \AA $^{3}$ \cite{LE14}. The binding experiments \cite{AM84} used
a fixed $C_{\text{Na}^{+}}^{\text{B}}=C_{\text{1}}^{\text{B}}=32$ mM and
various Ca$^{2+}$ bath concentrations $C_{\text{Ca}^{2+}}^{\text{B}%
}=C_{\text{2}}^{\text{B}}$ that imply different $O_{1}^{b}$ and $O_{2}^{b}$ of
Na$^{+}$ and Ca$^{2+}$ occupying the binding site. The occupancy numbers
$O_{1}^{b}$ and $O_{2}^{b}$ are determined by
\begin{equation}
\frac{O_{1}^{b}}{O_{2}^{b}}=\frac{1-O_{2}^{b}}{O_{2}^{b}}=\exp(-(\beta
_{1}-\beta_{2})\overline{\phi}_{b})\frac{C_{1}^{\text{B}}}{C_{2}^{\text{B}}},
\tag{22}%
\end{equation}
where $\overline{\phi}_{b}$ was just obtained from the case of equal
occupancy. The occupancy ratio in (22) thus deviates from unity as
$C_{2}^{\text{B}}$ is varied along the horizontal axis of the binding curve
from its midpoint value $C_{2}^{\text{B}}=0.9$ $\mu$M as shown in Fig. 5 in
\cite{LE14}.

Keeping $\overline{\phi}_{b}$ fixed is equivalent to assuming that the
relation (22) between the occupancy and bath concentration ratios is linear
\cite{LE14}. Moreover, keeping $\overline{\phi}_{b}$ fixed in (20) is
equivalent to assuming that the O$^{1/2-}$ ions ($c_{j}$) move continuously in
response to the continuous change of charges $O_{1}^{b}z_{\text{Na}^{+}}%
+O_{2}^{b}z_{\text{Ca}^{2+}}$ in the binding site. In \cite{LE14}, the charge
change from $z_{\text{Na}^{+}}$ (Na$^{+}$ occupying the site) to
$z_{\text{Ca}^{2+}}$ (Ca$^{2+}$ occupying the site) reflects a change of pore
radius of about 2.3 \AA \ that is surprisingly close to the value of 2
\AA \ obtained by MD simulations \cite{BG02}. Note that the vacuum
permittivity $\epsilon_{0}$ is chosen in (20) since both MD models in
\cite{LF01,BG02} treat O$^{1/2-}$ ions explicitly as shown in Fig. 1 (or Fig.
9 in \cite{LF01}). The Coulomb forces between the binding ion and O$^{1/2-}$
ions should therefore be calculated in vacuum since nothing is in between
these ions. Our numerical results can thus be verified with those of MD. Of
course, our assumptions in the linear model should be modified if more
accurate structural information can be used to provide an extra equation for
variable $\overline{\phi}_{b}$. The permittivity is chosen as 30$\epsilon_{0}$
\cite{LE14} in our forthcoming studies on the protein structure of a
sodium/calcium exchanger (NCX) \cite{LL12}. Meanwhile, the linear model seems
at least as good as the homology structure itself, and provides potentially
useful and interesting insights as we show in the results section.
Nevertheless, we imagine that nature might design flexible proteins so that
$\overline{\phi}_{b}$ is fixed or slightly perturbed by small thermal
variations so that the linear model is still tolerable within numerical errors
in theoretical simulations.

The simple atomic structure in Fig. 2 elucidates algebraic and subsequent PNPF
calculations in a concise way. The molecular-continuum model presented here
can be extended to deal with more complex nonequilibrium systems in real
protein channels in future studies. Application of the algebraic model to the
NCX structure \cite{LL12} is briefly discussed in \cite{LE14}. It will be
interesting to apply the present model to recent structures of a TRPV1 ion
channel \cite{LC13} and a voltage gated calcium channel \cite{TC14}.

For nonequilibrium cases, the binding steric potential $\overline{S}%
_{b}^{\text{trc}}$ is assigned its equilibrium value in subsequent PNPF
calculations, i.e., the void fraction $\Gamma(\mathbf{r})$ in $\overline
{\Omega}_{\text{Bind}}$ is assumed to remain unchanged from equilibrium to
nonequilibrium. The electrical potential $\overline{\phi}_{b}$ will be
modified by the membrane potential $V_{\text{i}}-V_{\text{o}}$ \cite{KC99} and
then used as a Dirichlet type condition for the potential function
$\phi(\mathbf{r})$ in $\overline{\Omega}_{\text{Bind}}$. We summarize the
boundary conditions for the PF (9) and NPF (15) equations defined in the
solvent region $\Omega_{s}$ in Fig. 1 as%
\begin{equation}
\left\{
\begin{array}
[c]{l}%
\text{ }\phi(\mathbf{r})=\widetilde{\phi}_{b}(\mathbf{r})\text{ in }%
\overline{\Omega}_{\text{Bind}}\text{, }\phi(\mathbf{r})=V_{\text{o,i}}\text{
on }\partial\Omega_{\text{Bath}}\text{,}\\
\nabla\phi(\mathbf{r})\cdot\mathbf{n}=0\text{ on }\partial\Omega_{s}%
\backslash\partial\Omega_{f}\text{,}\\
C_{1}(\mathbf{r})=C_{1}^{\text{B}}=\text{[Na}^{+}\text{]}_{\text{o,i}}\text{,
}C_{2}(\mathbf{r})=C_{2}^{\text{B}}=\text{[Ca}^{2+}\text{]}_{\text{o,i}%
}\text{, }C_{3}(\mathbf{r})=C_{3}^{\text{B}}=\text{[Cl}^{-}\text{]}%
_{\text{o,i}}\text{ on }\partial\Omega_{\text{Bath}}\text{,}\\
\mathbf{J}_{i}(\mathbf{r})\cdot\mathbf{n}=0\text{ on }\partial\Omega
_{s}\backslash\partial\Omega_{\text{Bath}}\text{, }%
\end{array}
\right.  \tag{23}%
\end{equation}
where $\mathbf{n}$ is an outward unit vector on the solvent boundary
$\partial\Omega_{s}$. Note that the electrostatic potential $\phi(\mathbf{r})$
is prescribed as a Dirichlet function $\widetilde{\phi}_{b}(\mathbf{r})$ whose
spatial average in $\overline{\Omega}_{\text{Bind}}$ is the constant
$\overline{\phi}_{b}$. However, the binding domain $\overline{\Omega
}_{\text{Bind}}$ is treated as an interior domain instead of boundary domain
for the flux equation (15). An iterative process of solving PF (9) and then
NPF (15) is repeated until self-consistent solutions of $\phi(\mathbf{r})$ and
$C_{i}(\mathbf{r})$ are reached within a tolerable error bound.

Treating the interior domain $\overline{\Omega}_{\text{Bind}}\subset$
$\Omega_{s}$ in place of the conventional boundary $\partial\Omega_{s}$ for
the potential condition $\phi(\mathbf{r})=\widetilde{\phi}_{b}(\mathbf{r})$ is
not a conventional way to solve the Poisson equation. This method is needed
because the binding potential $\overline{\phi}_{b}$ determined by (18)-(20) is
coupled to the steric potential $\overline{S}_{b}^{\text{trc}}$ that in turn
depends crucially on the conformation of the binding ion and protein atoms,
their interstitial voids, and their charges as shown in Fig. 2. Water also
plays a vital role as in (19). The steric equation (19) is in the interior
domain $\overline{\Omega}_{\text{Bind}}$. In other words, for calculating
$\overline{S}_{b}^{\text{trc}}$ we need to consider voids and water volume
that are interior quantities in $\overline{\Omega}_{\text{Bind}}$ and cannot
be specified on the solvent boundary $\partial\Omega_{s}$. We thus need to
impose $\phi(\mathbf{r})=\widetilde{\phi}_{b}(\mathbf{r})$ in $\overline
{\Omega}_{\text{Bind}}$ because $\widetilde{\phi}_{b}(\mathbf{r})$ is given by
$\overline{\phi}_{b}$ in $\overline{\Omega}_{\text{Bind}}$.

If a conventional method is used to solve the Poisson (or PF) equation
\cite{L13}, the resulting steric potential $\overline{S}_{b}^{\text{trc}}$ (as
an output of $\phi(\mathbf{r})$ by (7)) may be completely incorrect in
$\overline{\Omega}_{\text{Bind}}$ because the atomic equations (19) and (20)
are not used. We do not have any differential equation for the steric function
$S^{\text{trc}}(\mathbf{r})$ for which appropriate boundary conditions near
$\overline{\Omega}_{\text{Bind}}$ can be imposed if a conventional method is
used. Moreover, the important role of water and its volume effect is not taken
into account in conventional methods or models.

The models and methods proposed in this paper are still coarse approximations
to ion transport as the PNPF theory is in its early development. Nevertheless,
the theory proposed here provides many atomic properties such as (19) and (20)
that have been shown to be important for studying the binding mechanism in
Ca$_{\text{V}}$ channels \cite{LE14} and are also important for the transport
mechanism as shown in the next section. Incorporating atomic properties into
continuum models is a step forward to improve and refine the continuum theory
that has been challenged for its accuracy when compared to (mostly equilibrium
calculations) MC, BD, or MD \cite{SM03,CK00,IR01,IR02}. Continuum models on
the other hand have substantial advantages in efficiency that are of great
importance in studying a range of conditions and concentrations, as are
present in experiments.

\section{Results}

Ca$_{\text{V}}$ channel conducts primarily Na$^{+}$ when the Ca$^{2+}$
concentration is below 1 $\mu$M but it conducts primarily Ca$^{2+\text{ }}$ in
the physiological concentration range mM. In \cite{AM84}, 19 extracellular
solutions and 3 intracellular solutions were studied experimentally. The range
of [Ca$^{2+}$]$_{\text{o}}$ is 10$^{8}$-fold from $10^{-10.3}$ to $10^{-2}$ M
as given in \cite{AM84}. Explaining the biological function of trace Ca$^{2+}$
concentrations is a crucial task of biophysical models while dealing with the
large Ca$^{2+}$ concentrations found in extracellular solutions of all
biological systems. This range of calcium concentrations poses severe
obstacles for MD and BD simulations even on the most advanced computers to
date \cite{E10}. To our knowledge, a comparison of MD simulations with
experimental measurements \cite{AM84} has not yet been reported without
invoking arbitrary (i.e., untested) extrapolation methods for handling the
10$^{8}$-fold variation of concentration and the dynamics of ionic flow
\cite{E10,BN08}.

PNPF results are in accord with the experimental data in \cite{AM84} as shown
in Fig. 3 under only the same salt conditions of NaCl and CaCl$_{2}$ in pure
water, without considering protons and other bulk salts in experimental
solutions. With [$\text{Na}^{+}$]$_{\text{i}}=$ [$\text{Na}^{+}$]$_{\text{o}%
}=32$ mM and [Ca$^{2+}$]$_{\text{i}}=0$, the membrane potential is fixed at
$-20$ mV ($V_{\text{o}}=0$ and $V_{\text{i}}$ $=-20$ mV) throughout, as
assumed in Fig. 11 of \cite{AM84} for all single-channel currents (in femto
ampere fA) recorded in the experiment. Note that the experimental currents
have been converted to single-channel currents in Fig. 11 of \cite{AM84} as
shown on the right hand $y$-axis in that figure. The currents are on the
femtoscale because calcium channels have been long recognized to be in some
sense blocked sodium channels and here too we have small membrane potentials.
The small circles in Fig. 3 denote the estimated currents (by eye) from Fig.
11 of \cite{AM84} and the plus sign denotes the current calculated by PNPF.
The channel is almost blocked by Ca$^{2+}$ ions with a current of about 15 fA
at [Ca$^{2+}$]$_{\text{o}}=10^{-4.2}$ M. Half-blockage current (about 77 fA)
is defined by the one half of the saturation current (about 154 fA at
[Ca$^{2+}$]$_{\text{o}}=10^{-10.3}$ M). The half-blockage Ca$^{2+}$
concentration is about [Ca$^{2+}$]$_{\text{o}}=0.9$ $\mu$M and that is used to
define the midpoint binding condition (21).

PNPF deals naturally with the main experimental data of ionic flow based on
this binding definition. Physical parameters in (9) and (16) are summarized in
Table 1 with their physical meaning, numerical values, and units for ease of
reference. The diffusion coefficient in the channel pore is taken as
$\theta_{i}D_{i}$ for each ionic species, where $D_{i}$ are bulk values in the
table and $\theta_{i}=0.1$ are factors for the pore. All physical parameters
are kept fixed throughout.

PNPF can provide more physical details of ion transport inside the channel
pore such as electrical potential ($\phi(\mathbf{r})$ in Fig. 4), steric
potential ($S^{\text{trc}}(\mathbf{r})$ in Fig. 5), energy wells
($\mathcal{E}_{2}(\mathbf{r})$ in Fig. 6), water density ($C_{4}(\mathbf{r})$
shown in Fig. 7), void volume fraction ($\Gamma(\mathbf{r})$ in Fig. 8),
dielectric function ($\widetilde{\epsilon}(\mathbf{r})$ in Fig. 9), crowded
ions in binding site (Fig. 10), concentrations ($C_{2}(\mathbf{r})$ in Fig.
11), and flux densities ($\left\vert \mathbf{J}_{2}(\mathbf{r})\right\vert $
in Fig. 12). The electric potential profiles remain almost unchanged for
various $C_{2}^{\text{B}}$ as shown in Fig. 4 following the linear model of
the occupancy equation (22).

\textbf{\underline{\textbf{What is new?}} }The steric potential profiles shown
in Fig. 5 represent the novelty of the PNPF theory. All effects of volume
exclusion, interstitial void, configuration entropy, short range interactions,
correlation, polarization, screening, and dielectric response of this nonideal
system are described by the steric functional $S^{\text{trc}}(\mathbf{r})$.
The Ca$^{2+}$ energy landscape $\mathcal{E}_{2}(\mathbf{r})=(\beta_{2}%
\phi(\mathbf{r})-S^{\text{trc}}(\mathbf{r}))k_{B}T$ (Fig. 6) is modified by
this potential. The steric potential in the binding region decreases
drastically from $-1.30$ to $-10.34$ $k_{B}T$ as $C_{2}^{\text{B}}$ increases
from $10^{-7.2}$ to $10^{-2}$ M. The water density $C_{4}(\mathbf{r}%
)=C_{4}^{\text{B}}\exp(S^{\text{trc}}(\mathbf{r}))$ (Fig. 7) and the void
volume $\Gamma(\mathbf{r)}$ (Fig. 8) decrease as well because more Ca$^{2+}$
ions in the bath make the binding site more crowded as was previously seen in
the algebraic Fermi model \cite{LE14}.%
\begin{figure}
[ptb]
\begin{center}
\includegraphics[
height=2.6991in,
width=4.5991in
]%
{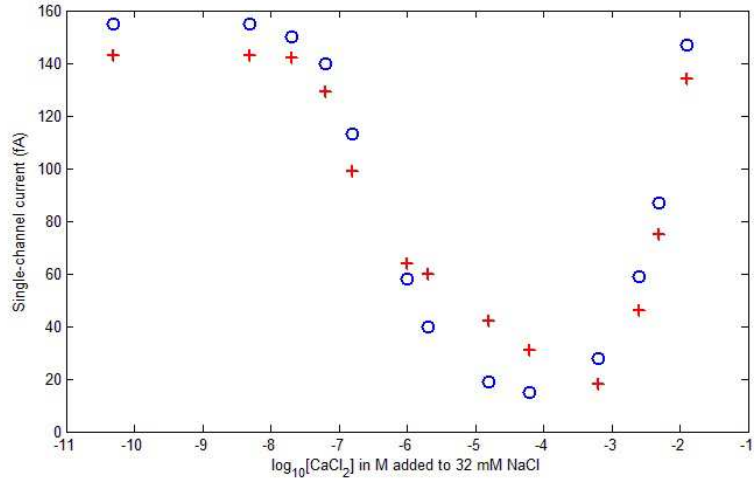}%
\caption{Anomalous Mole Fraction Effect. Single-channel inward current in
femto ampere (fA) plotted as a function of $\log_{10}$[Ca$^{2+}$]$_{\text{o}}%
$. Experimental data marked by small circles are those in \cite{AM84} whereas
the PNPF data are denoted by the plus sign.}%
\end{center}
\end{figure}

\begin{center}
$%
\begin{tabular}
[c]{c|c|c|c}%
\multicolumn{4}{c}{Table 1. Notations and Physical Constants}\\\hline
Symbol & Meaning & \ Value & \ Unit\\\hline
\multicolumn{1}{l|}{$k_{B}$} & \multicolumn{1}{|l|}{Boltzmann constant} &
\multicolumn{1}{|l|}{$1.38\times10^{-23}$} & \multicolumn{1}{|l}{J/K}\\
\multicolumn{1}{l|}{$T$} & \multicolumn{1}{|l|}{temperature} &
\multicolumn{1}{|l|}{$298.15$} & \multicolumn{1}{|l}{K}\\
\multicolumn{1}{l|}{$e$} & \multicolumn{1}{|l|}{proton charge} &
\multicolumn{1}{|l|}{$1.602\times10^{-19}$} & \multicolumn{1}{|l}{C}\\
\multicolumn{1}{l|}{$\epsilon_{0}$} & \multicolumn{1}{|l|}{permittivity of
vacuum} & \multicolumn{1}{|l|}{$8.85\times10^{-14}$} &
\multicolumn{1}{|l}{F/cm}\\
\multicolumn{1}{l|}{$\epsilon_{\text{w}}$} & \multicolumn{1}{|l|}{water
dielectric constant} & \multicolumn{1}{|l|}{$78.5$} & \multicolumn{1}{|l}{}\\
\multicolumn{1}{l|}{$l_{c}$} & \multicolumn{1}{|l|}{correlation length} &
\multicolumn{1}{|l|}{$1.98$} & \multicolumn{1}{|l}{\AA }\\
\multicolumn{1}{l|}{$D_{1}$} & \multicolumn{1}{|l|}{Na$^{+}$ diffusion
coefficient} & \multicolumn{1}{|l|}{$1.334\times10^{-5}$} &
\multicolumn{1}{|l}{cm$^{2}/$s}\\
\multicolumn{1}{l|}{$D_{2}$} & \multicolumn{1}{|l|}{$\text{Ca}^{2+}$ diffusion
coefficient} & \multicolumn{1}{|l|}{$0.792\times10^{-5}$} &
\multicolumn{1}{|l}{cm$^{2}/$s}\\
\multicolumn{1}{l|}{$D_{3}$} & \multicolumn{1}{|l|}{$\text{Cl}^{-}$ diffusion
coefficient} & \multicolumn{1}{|l|}{$2.032\times10^{-5}$} &
\multicolumn{1}{|l}{cm$^{2}/$s}\\
\multicolumn{1}{l|}{$C_{1}^{\text{B}}$} & \multicolumn{1}{|l|}{Na$^{+}$ bath
concentration} & \multicolumn{1}{|l|}{$32$} & \multicolumn{1}{|l}{mM}\\
\multicolumn{1}{l|}{$C_{2}^{\text{B}}$} & \multicolumn{1}{|l|}{$\text{Ca}%
^{2+}$ bath concentration} & \multicolumn{1}{|l|}{$10^{-10.3}$ $\sim$
$10^{-2}$} & \multicolumn{1}{|l}{M}\\
\multicolumn{1}{l|}{$V_{\text{i,o}}$} & \multicolumn{1}{|l|}{inside (outside)
voltage} & \multicolumn{1}{|l|}{$0$ ($-20$)} & \multicolumn{1}{|l}{mV}\\\hline
\end{tabular}
\ $
\begin{figure}
[ptb]
\begin{center}
\includegraphics[
height=2.7086in,
width=4.6164in
]%
{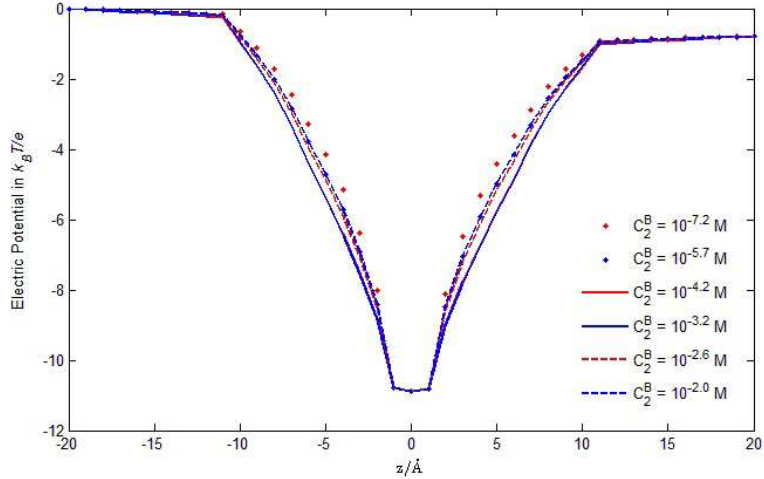}%
\caption{The electrical potential $\phi(\mathbf{r})$ profiles (averaged over
each cross section) along the pore axis for various $C_{2}^{\text{B}}$ ranging
from 10$^{-7.2}$ M to 10$^{-2}$ M. All the following figures are obtained with
the same averaging method and the same range of $C_{2}^{\text{B}}$.}%
\end{center}
\end{figure}
\begin{figure}
[ptbptb]
\begin{center}
\includegraphics[
height=2.7086in,
width=4.6164in
]%
{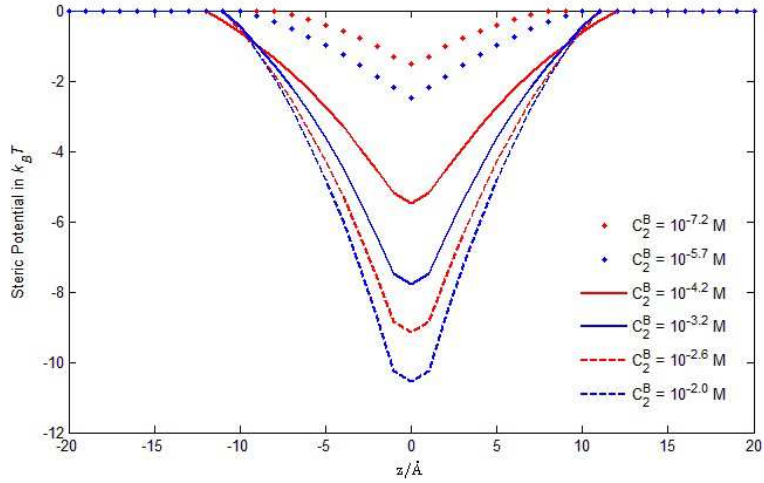}%
\caption{The averaged steric potential $S^{\text{trc}}(\mathbf{r})$ profiles.}%
\end{center}
\end{figure}

\end{center}

In physiological bath conditions $C_{2}^{\text{B}}=10^{-2}\sim10^{-3}$ M, Fig.
7 shows that the region containing the binding site with the length about 10
\AA \ is very dry (hydrophobic), which agrees with the recent crystallographic
analysis \cite{LL12} of the Ca$^{2+}$ binding site of the related protein NCX
with the EETT locus showing a hydrophobic patch (also about 10 \AA \ in
length) formed by the conserved Pro residues. The hydrophobicity near the
binding site in our model is described by the continuous water density
function $C_{4}(\mathbf{r})$ via the continuous steric function $S^{\text{trc}%
}(\mathbf{r})$ as shown in Fig. 5. At $C_{2}^{\text{B}}=10^{-2}$ M, the
magnitude of the steric energy $S^{\text{trc}}=-10.34$ $k_{B}T$ is comparable
to that of the electrostatic energy $\phi=-10.48$ $k_{B}T/e$. This
surprisingly large energy due only to the steric effect has not been
quantified and observed by MD, MC, or other continuum methods in
Ca$_{\text{V}}$ channel modeling, as far as we know. As observed from Fig. 3,
ionic transport is blocked by the competition between Na$^{+}$ and Ca$^{2+}$
ions in the range $C_{2}^{\text{B}}=10^{-5.7}\sim10^{-4.2}$ M. In this
blocking range, the corresponding steric profiles in Fig. 5 are wider
indicating that the water density or the void volume is more evenly distributed.

Fig. 9 demonstrates the combined effects of correlation, polarization, and
screening in this highly inhomogeneous electrolyte by means of the variation
of dielectric coefficient produced by the Poisson-Fermi equation (9). Note
that the dielectric function can also be calculated by $\widetilde{\epsilon
}(\mathbf{r})=\epsilon_{b}+C_{4}(\mathbf{r})(\epsilon_{\text{w}}-\epsilon
_{b})/C_{4}^{\text{B}}$ using the water density function $C_{4}(\mathbf{r})$
as proposed in \cite{LM08}, where $\epsilon_{b}=2$. Fig. 10 illustrates how
ions crowd in the highly charged binding site and, in the meantime, voids and
water vacate under the condition $C_{2}^{\text{B}}=10^{-5.7}$ M.
\begin{figure}
[ptb]
\begin{center}
\includegraphics[
height=2.7034in,
width=4.6077in
]%
{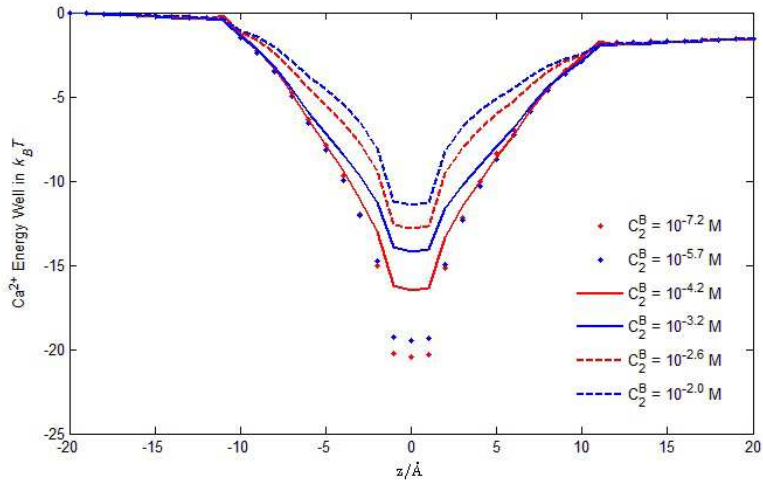}%
\caption{The averaged Ca$^{2+}$ energy well $\mathcal{E}_{2}(\mathbf{r})$
profiles.}%
\end{center}
\end{figure}
\begin{figure}
[ptbptb]
\begin{center}
\includegraphics[
height=2.7034in,
width=4.6077in
]%
{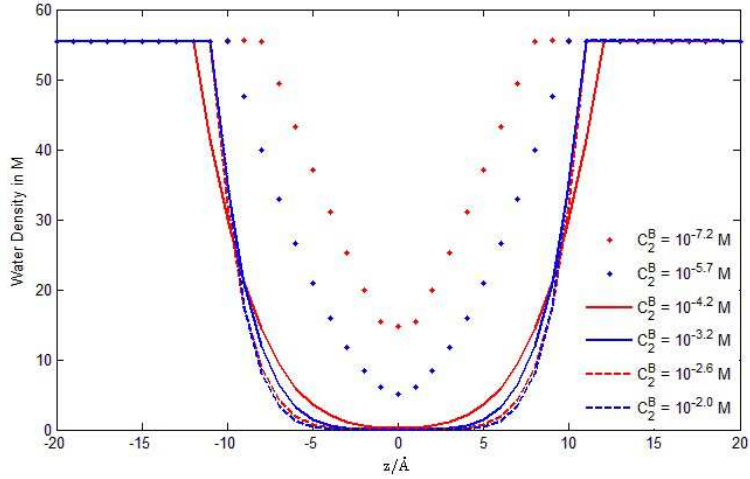}%
\caption{The averaged water density $C_{4}(\mathbf{r})$ profiles.}%
\end{center}
\end{figure}
\begin{figure}
[ptbptbptb]
\begin{center}
\includegraphics[
height=2.7086in,
width=4.6164in
]%
{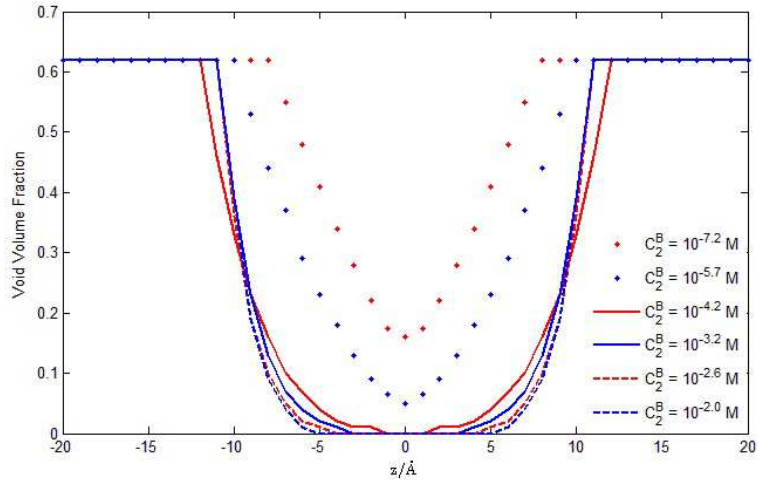}%
\caption{The averaged void volume fraction $\Gamma(\mathbf{r})$ profiles.}%
\end{center}
\end{figure}
\begin{figure}
[ptbptbptbptb]
\begin{center}
\includegraphics[
height=2.6991in,
width=4.5991in
]%
{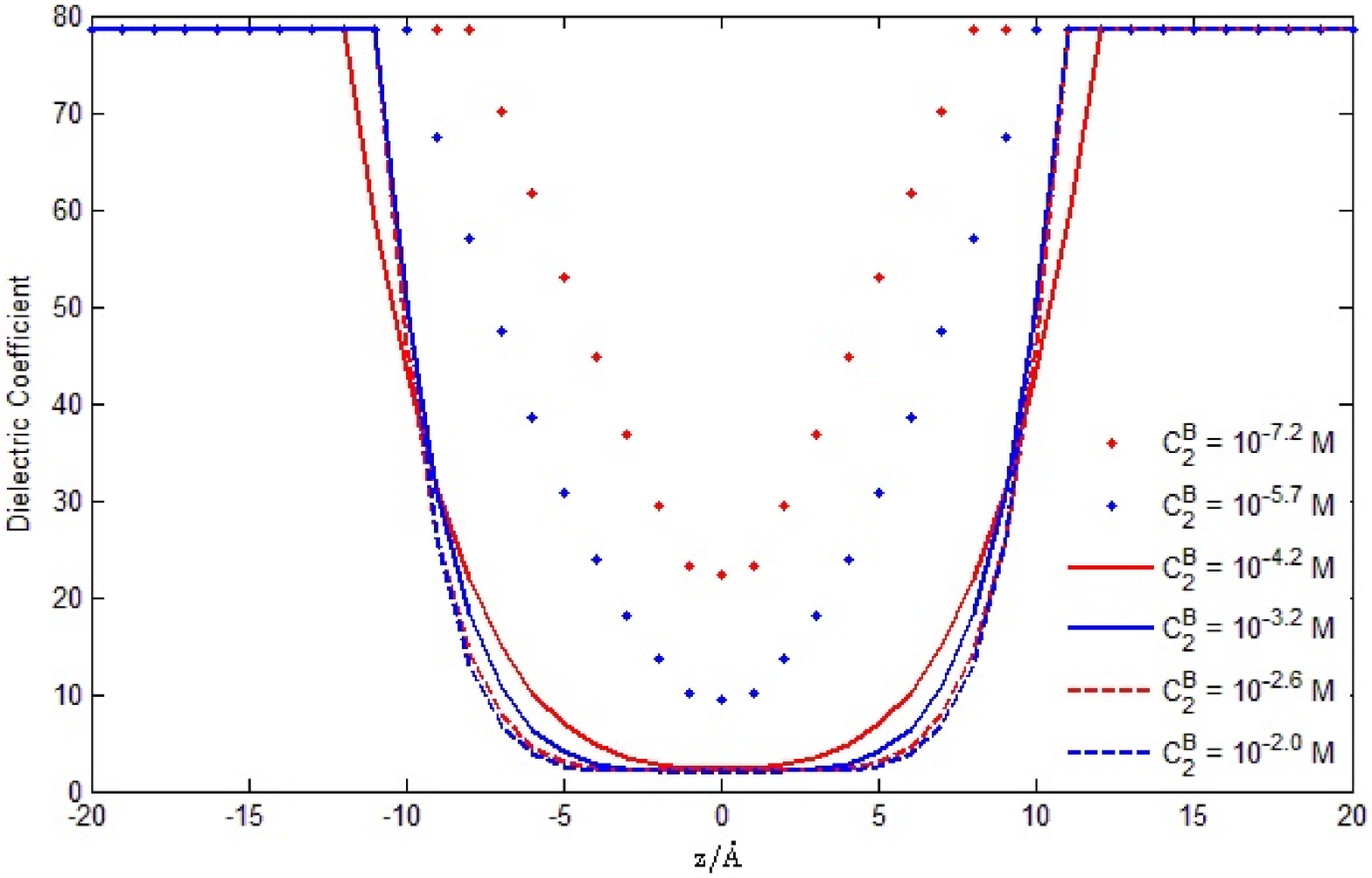}%
\caption{The averaged dielectric function $\widetilde{\epsilon}(\mathbf{r})$
profiles.}%
\end{center}
\end{figure}
\begin{figure}
[ptbptbptbptbptb]
\begin{center}
\includegraphics[
height=2.7086in,
width=4.6164in
]%
{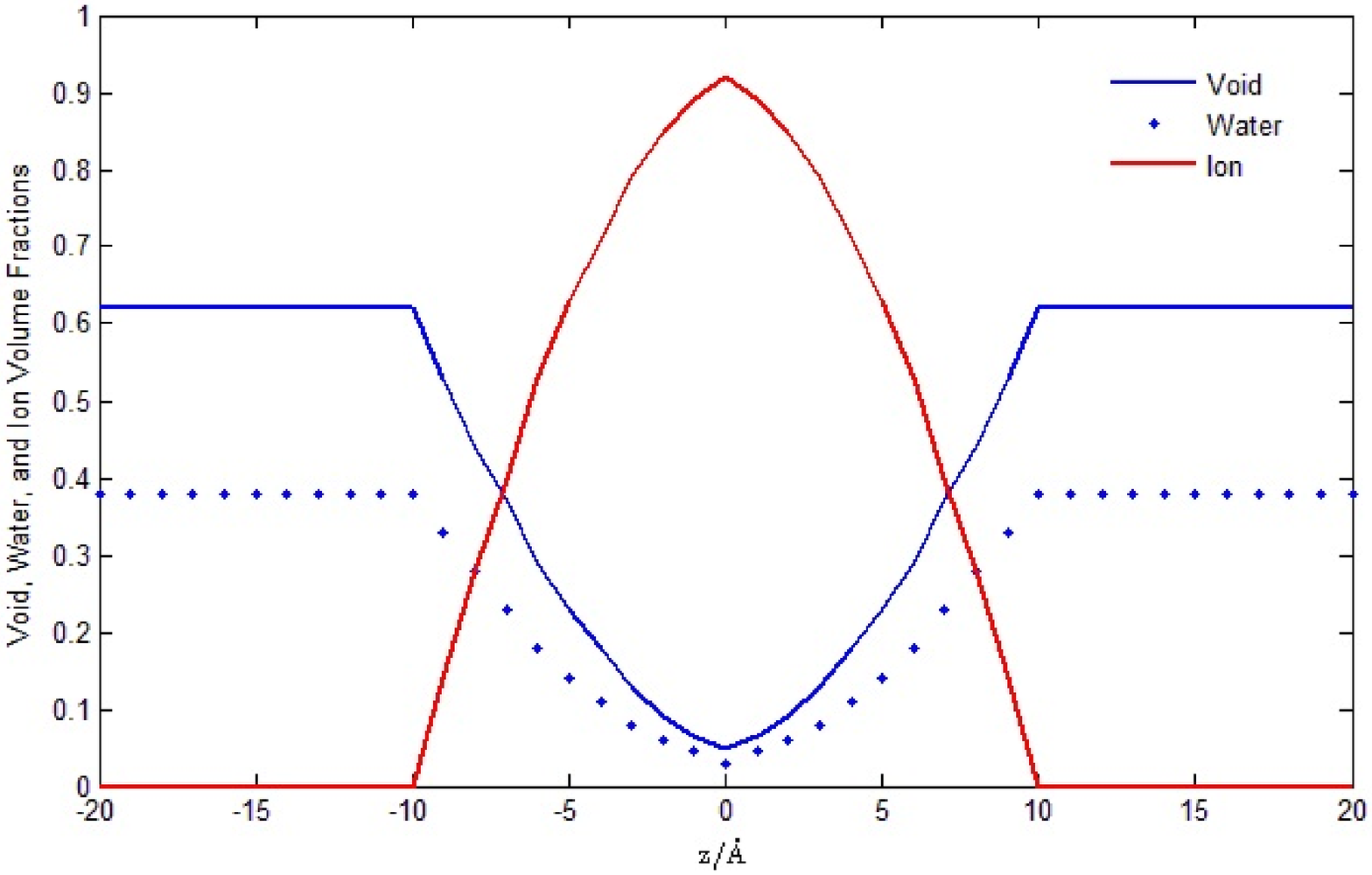}%
\caption{Crowded Charge in Binding Site. The volume fractions of voids, water,
and ions per unit volume at the Ca$^{2+}$ bath concentration $C_{2}^{\text{B}%
}=10^{-5.7}$ M.}%
\end{center}
\end{figure}
\begin{figure}
[ptbptbptbptbptbptb]
\begin{center}
\includegraphics[
height=2.7086in,
width=4.6164in
]%
{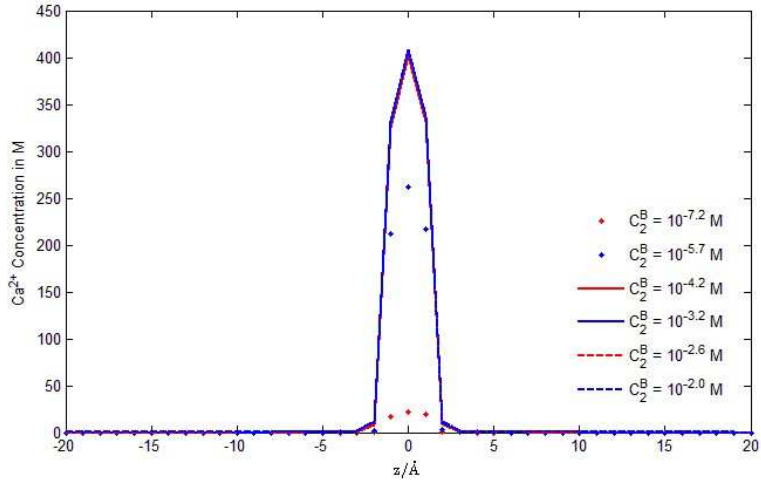}%
\caption{The averaged Ca$^{2+}$ concentration $C_{2}(\mathbf{r})$ profiles.}%
\end{center}
\end{figure}
\begin{figure}
[ptbptbptbptbptbptbptb]
\begin{center}
\includegraphics[
height=2.7294in,
width=4.651in
]%
{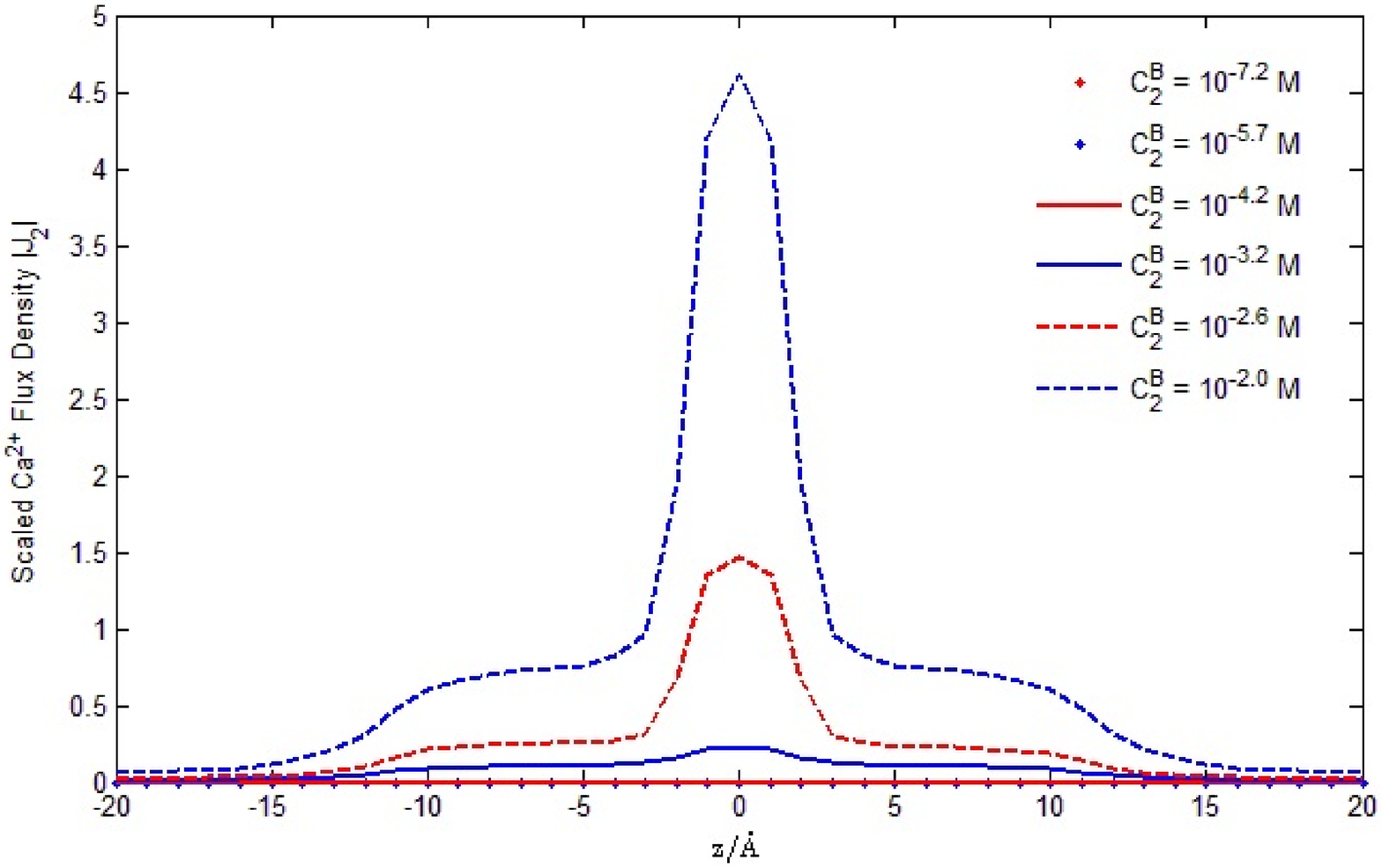}%
\caption{The averaged Ca$^{2+}$ flux density $\left\vert \mathbf{J}%
_{2}(\mathbf{r})\right\vert $ profiles.}%
\end{center}
\end{figure}

The Ca$^{2+}$ occupancy $O_{2}^{b}$ in the binding site increases from $0.69$
(not shown) at $C_{2}^{\text{B}}=10^{-5.7}$ M to almost 1 at $C_{2}^{\text{B}%
}=10^{-2}$ M. The corresponding peak Ca$^{2+}$ concentration shown in Fig. 11
also increases from 261.83 to 408.53 M in this range. The largest
concentration is still below the maximum allowable value $C_{2}^{\text{Max}%
}=1/v_{2}=408.57$ M as implied by the Fermi distribution (7). We obtained this
incredibly large concentration since $O_{2}^{b}\approx1$ in (19), which in
turn yields $1/v_{b}\approx408.53$ M as $v_{b}\approx v_{2}$ indicating the
importance of the atomic nature of the steric potential (19) with variable
$v_{b}$. This figure demonstrates that the PNPF model can capture the atomic
properties of ions in flow, a critical (however small) step in continuum
theory toward the ultimate accuracy in theoretical simulations. From Fig. 3,
we observe that Ca$^{2+}$ currents increase dramatically when [Ca$^{2+}%
$]$_{\text{o}}$ increases from $10^{-3.2}$ to $10^{-2}$ M in the physiological
mM range of Ca$_{\text{V}}$ channels. The corresponding flux density profiles
in the binding region increase dramatically too as shown in Fig. 12.

We make a final remark about these results. As intensively studied in
\cite{CK00,IR01,IR02}, the classical PNP model fails to yield ionic
concentrations and currents properly (compared with those of Brownian
dynamics), especially for narrow channels, because the classical model does
not include the volume effect of ions, water, and voids, the correlation
effect of ions, and the screening effect of water. The PNPF model not only
computes ionic currents comparable to the experimental data (Fig. 3) for the
narrow calcium channel but also provides many physical properties (Figs. 5-10,
for example) that are shown to depend critically on the proposed steric
potential. The PNPF model overcomes the limitation of the classical PNP model
with respect to these effects and properties.

\section{Conclusion}

We propose a Poisson-Nernst-Planck-Fermi model for studying equilibrium and
nonequilibrium systems of ionic liquids and solution electrolytes. The
excluded volume effect of different sizes of ions and water molecules, the
correlation effect of crowded ions, and the screening effect of polar water
molecules in inhomogeneous aqueous electrolytes are all included in this
model. The model was verified by a set of experimental currents of L-type Ca
channels recorded in a 10$^{8}$-fold range of Ca$^{2+}$ concentrations that
show the exceptional selectivity of these channels.

We also propose a consistent Gibbs free energy functional leading to a Fermi
like distribution of hard spherical particles in the electrolytic system. The
Gibbs-Fermi functional is shown to converge to the Gibbs-Boltzmann functional
that yields the Boltzmann distribution as the volumes of all particles and the
correlation length approach zero. Moreover, we introduce a Gibbs-Fermi entropy
for which the excluded volume of water molecules and the dynamic distribution
of interstitial voids between particles are needed to establish a consistent
generalization of nonideal inhomogeneous electrolytes for both equilibrium and
nonequilibrium systems. The local probability of any particle species in the
Gibbs-Fermi entropy is expressed in terms of the local and bulk ratios between
the particle and void fractions per unit volume. We show that if all particles
are treated as hard spheres, the voids must be included in the Gibbs free
energy functional. The voids are created by packing defects and Lennard-Jones
and Coulomb forces between particles. The void effect plays an essential role
in our theory as well as in our results, especially in the steric energy.

Most of the results in this article seem to be novel, because consistent
models including voids, water volume, and Fermi distributions have not been
developed previously, as far as we know. These numerical results provide
useful tools to develop insight into a variety of physical mechanisms ranging
from binding, to permeation, blocking, flexibility, and charge/space
competition of the channel.

\begin{acknowledgments}
This work was support in part by National Science Council of Taiwan under
Grant No. 102-2115-M-134-005 to J.L.L. and by the Bard Endowed Chair of Rush
University Medical Center, held by B.E.
\end{acknowledgments}

\end{document}